\documentclass[a4paper,preprint,byrevtex,showpacs,aip]{revtex4-1}

\DeclareSymbolFont{lettersA}{U}{pxmia}{m}{it}
\DeclareMathAlphabet{\mathsfsl}{OT1}{cmss}{m}{sl}

\SetSymbolFont{lettersA}{bold}{U}{pxmia}{bx}{it}
\DeclareFontSubstitution{U}{pxmia}{m}{it}
\DeclareSymbolFontAlphabet{\mathfrak}{lettersA}
\DeclareMathSymbol{\piup}{\mathord}{lettersA}{"19}
\DeclareMathSymbol{\iTheta}{\mathalpha}{letters}{2}

\usepackage[pagewise]{lineno}
\usepackage{amsmath}
\usepackage{amssymb}
\usepackage{graphicx}
\usepackage{epstopdf}
\usepackage[hang,scriptsize]{subfigure}
\usepackage{bbm}
\usepackage{mathrsfs}
\usepackage{xcolor}
\usepackage[dvipdfm,colorlinks=true,pdfstartview=FitV,linkcolor=blue,citecolor=blue,urlcolor=blue]{hyperref}

\makeatletter
\newcommand{\rmnum}[1]{\romannumeral #1}
\newcommand{\Rmnum}[1]{\expandafter\@slowromancap\romannumeral #1@}
\makeatother

\newcommand{\ii}{\mathrm{i}}

\newcommand{\blue}[1]{\textcolor{blue}{#1}}

\begin{document}

% If you do not want to show the line numbers %
% Just comment the "\linenumbers" %
%\linenumbers

% Use the following commands to number the equations %
% \begin{linenomath} %
% \begin{equation} %
% %% %
% \end{equation} %
% \end{linenomath} %
%
% Use the following commands to omit the equations %
% \begin{linenomath*} %
% \begin{equation} %
% %% %
% \end{equation} %
% \end{linenomath*} %

\title{Strategy for designing broadband
    epsilon-near-zero metamaterial
    with loss compensation by gain media}

\author{L. Sun}
\email{lsun@phy.cuhk.edu.hk}
\affiliation{Department of Physics, \\
    The Chinese University of Hong Kong, \\
    Shatin, N.T., Hong Kong}

\author{K. W. Yu}
\email{kwyu@phy.cuhk.edu.hk}
\affiliation{Department of Physics, \\
    The Chinese University of Hong Kong, \\
    Shatin, N.T., Hong Kong}

\date{\today}

\begin{abstract}
A strategy is proposed to design the broadband gain-doped
epsilon-near-zero (GENZ) metamaterial.
Based on the Milton representation of effective permittivity,
the strategy starts in a dimensionless spectral space,
where the effective permittivity of GENZ metamaterial is simply
determined by a pole-zero structure corresponding to the
operating frequency range.
The physical structure of GENZ metamaterial is retrieved from
the pole-zero structure via a tractable inverse problem.
The strategy is of great advantage in practical applications
and also theoretically reveals the cancellation mechanism
dominating the broadband near-zero permittivity phenomenon
in the spectral space.
\end{abstract}

%\pacs{TO BE DONE...}

\maketitle

%%%%%%%%%%%%%%%%%%%%%%%%
%\section{Introduction}%
%%%%%%%%%%%%%%%%%%%%%%%%

Materials with near-zero permittivity (ENZ) have been vastly explored
due to their anomalous electromagnetic properties and unique applications,
such as
directive antenna \cite{Enoch2002PRL},
subwavelength imaging system \cite{Salandrino2006PRB, Silveirinha2009PRL},
optical nanocircuit \cite{Engheta2007SCI},
radiation phase pattern converter \cite{Alu2007PRB},
reflectionless waveguide \cite{Silveirinha2007PRB1, Edwards2009JAP},
field energy confinement \cite{Silveirinha2007PRB2},
and electromagnetic transparency and cloaking design \cite{Alu2005PRE, Sun2011JOSAB}.
In nature, such materials are already available as noble metals and polar dielectrics.
Characterized by the Drude or Drude-Lorenz model \cite{Bohren1983BOOK, Jackson1999BOOK},
the real part of their permittivity approaches to zero near the plasma frequency.
However, the broadband ENZ material is required regarding to the practical applications.
To manufacture the broadband ENZ materal,
the basic idea is to make the metal-dielectric composite according to the
effective medium theory \cite{Goncharenkoa2010JN}.
But such mixture possesses a high loss in optical frequency range
due to the Kramers-Kronig relations \cite{John1956PR},
which limits the application scenario.
To overcome such disadvantage,
gain sources are suggested to reduce the loss \cite{Campione2011OME},
and due to the strong local field enhancement in ENZ material,
the gain sources can even provide large effective gain than when used alone
\cite{Bergman2003PRL, Stockman2008NP}.

Based on former study about the graded metal-dielectric materials
\cite{Huang2004APL, Dong2005PRE, Huang2006PR},
we have promoted a strategy to design an anisotropic broadband ENZ metamaterial
by applying the Bergman representation of effective permittivity \cite{Bergman1978PR, Bergman1992SSP}
in our previous work \cite{Sun2012JOSAB}.
However, a high loss exists in the operating frequency range due to the absence of gain sources.
In addition, the previous strategy requires analytical expressions of permittivities
of different components to determine the physical structure of the broadband ENZ metamaterial.
That limits its practical applications, where permittivities are usually numerically given~\cite{Johnson1972PRB}.
Therefore, we promote another strategy to design the anisotropic broadband gain-doped
epsilon-near-zero (GENZ) metamaterial base on Milton representation of effective permittivity
\cite{Milton1980APL, Milton1981JAP1, Milton1981JAP2} in this work.

The strategy is carried out in a dimensionless spectral space,
in which the effective permittivity of broadband GENZ metamaterial
is characterized by a series of pole-zero pairs,
initially arranged under lossless condition with respect to a properly
determined operating frequency range.
Then with proper variations on the initial zeros (zero-variation, for short),
the real part of effective permittivity can be near-zero in the operating frequency
range when loss is taken into account.
Finally, via a tractable inverse problem,
the physical structure of broadband GENZ metamaterial can be achieved
from the varied zeros and poles.
In contrast to the previous one,
this strategy can be carried out numerically with respect to the experimental data,
and due to the gain sources, the loss can be effectively reduced.
In the following, the strategy will be described in detail,
and it is worth noting that in order to simplify the description,
permittivities are still characterized by analytical formulae,
which is not necessary.

%%%%%%%%%%%%%%%%%%%%%%%%%%%%%
%\section{Structure Profile}%
%%%%%%%%%%%%%%%%%%%%%%%%%%%%%

\begin{figure}[htbp]
    \centering
    \includegraphics[width=8.5cm]{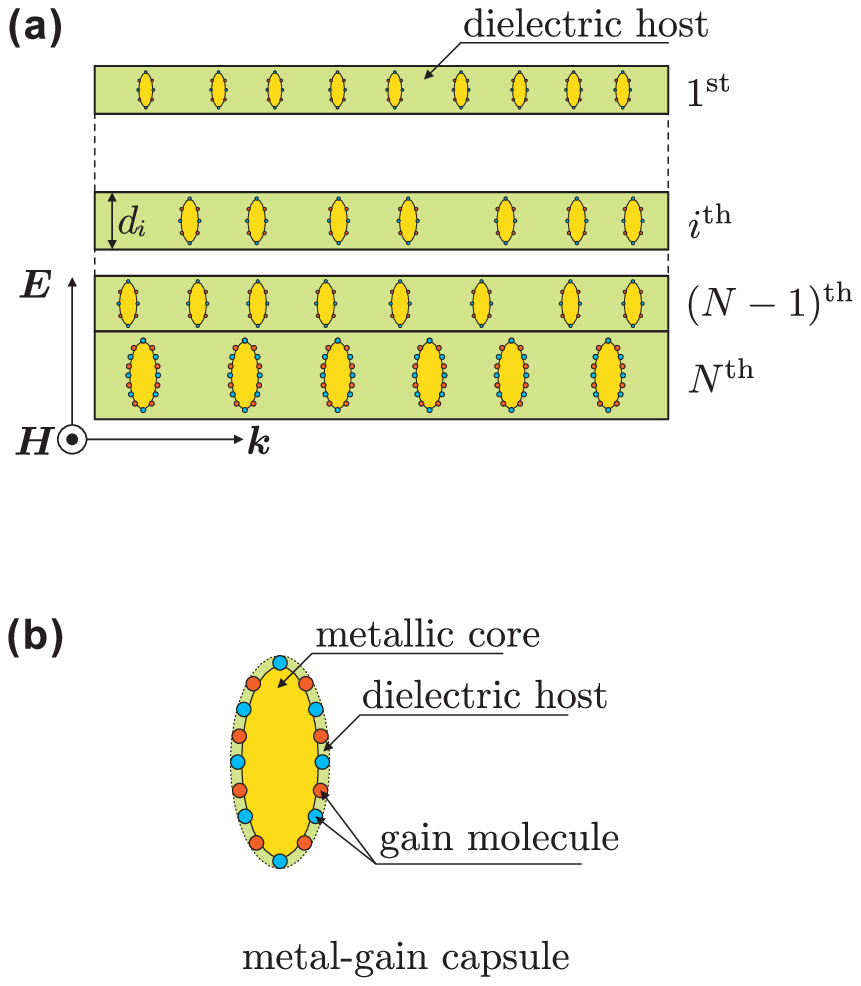}
    \caption{L.Sun and K.W.Yu}
    \label{fig:fig1}
\end{figure}

Depicted in Fig.~\ref{fig:fig1}, the broadband GENZ metamaterial is a flat $N$-layer stack.
Each layer is composed by two components: metal-gain capsules doped in a dielectric host.
%%%%-------------------------%%%
%\subsection{Component Profile}%
%%%%-------------------------%%%
The metal-gain capsule is a metallic core covered by a gain shell.
The permittivity of metallic core follows the Drude model
\begin{equation}
\label{eq:metal}
    \varepsilon_{m}(\omega) =
        \varepsilon_{\scriptscriptstyle\infty} - \frac{1}{\omega(\omega + \ii\gamma)},
\end{equation}
where $\varepsilon_{\scriptscriptstyle\infty}$ is a high frequency permittivity determined
to match the experimental data in visible region, and $\gamma$ is the damping factor.
The gain shell is composed by two kinds of gain molecules doped in dielectric host,
with an effective permittivity described by the Lorentz model~\cite{Campione2011OME}
\begin{equation}
\label{eq:gain}
    \varepsilon_{g}(\omega) = \varepsilon_{r}
        + \frac{\kappa_{1}}{\omega^{2} - \omega_{01}^{2} + \ii\sigma_{1}\omega}
        + \frac{\kappa_{2}}{\omega^{2} - \omega_{02}^{2} + \ii\sigma_{2}\omega},
\end{equation}
where $\varepsilon_{r}$ is the permittivity of dielectric host,
($\kappa_{1},\,\kappa_{2}$) are the density factors with respect to the number density of each gain molecule,
and ($\omega_{01},\,\omega_{02}$) are the center emission frequencies of each gain molecules
with the emission frequency line-widthes ($\sigma_{1},\,\sigma_{2}$).
For convenience, frequencies ($\omega,\,\omega_{01},\,\omega_{02}$), damping factor $\gamma$, and line-widthes
($\sigma_{1}\,\,\sigma_{2}$) are all normalized by the plasma frequency $\omega_{p}$ of metallic core.
Regarding the simple mixing rule, the effective permittivity of metal-gain capsule
can be described as
\begin{equation}
\label{eq:gain-media}
    \varepsilon_{1}(\omega) = p\,\varepsilon_{g}(\omega) + (1-p)\varepsilon_{m}(\omega),
\end{equation}
where $p$ is the volume fraction of gain shell.

For simplification, the permittivity of dielectric host of each layer is set to be
\begin{equation}
\label{eq:host}
    \varepsilon_{2} = \varepsilon_{r}.
\end{equation}
Besides, the thickness of $i$th-layer is denoted as $d_{i}$
and normalized by the total thickness of GENZ metamaterial.
Therefore, the following summation rule
\begin{equation}
\label{eq:di-sum}
    \sum_{i=1}^{N}d_{i} \equiv 1
\end{equation}
is held.

%%%%------------------------------%%%
%\subsection{Effective Permittivity}%
%%%%------------------------------%%%

For simplification, the effective permittivity of $i$th layer is described
by the simple mixing rule, as
\begin{equation}
\label{eq:i-layer}
    \varepsilon_{e}^{(i)}(\omega) = f_{i}\,\varepsilon_{1}(\omega) + (1 - f_{i})\varepsilon_{2},
\end{equation}
where $f_{i}$ is the volume fraction of metal-gain capsules in the $i$th layer.
By introducing a $s$-parameter
\begin{equation}
\label{eq:s-paramter}
    s = s(\omega) = \frac{\varepsilon_{2}}{\varepsilon_{2} - \varepsilon_{1}(\omega)},
\end{equation}
Eq.~\eqref{eq:i-layer} can be rewritten as a function of $s$-parameter
\begin{equation}
\label{eq:i-layer-s}
    \varepsilon_{e}^{(i)}(s) = \varepsilon_{2}\left(1 - \frac{f_{i}}{s}\right).
\end{equation}
Therefore, the effective permittivity of GENZ metamaterial is
\begin{equation}
\label{eq:stack}
    \varepsilon_{e}(s) = \varepsilon_{2}\left[\sum_{i=1}^{N}\frac{d_{i}}
        {1 - f_{i}/s}\right]^{-1}.
\end{equation}
According to this equation, a series of proper volume fractions ($f_{i}$) and thicknesses ($d_{i}$)
can make the effective permittivity of GENZ metamaterial, in the normal direction of the stack,
be near-zero in a wide frequency range.

On the other hand, the effective permittivity of GENZ metamaterial can also be described
by the Milton representation as
\begin{equation}
\label{eq:milton}
    \varepsilon_{e}(s) = \varepsilon_{2}
        \prod_{i=1}^{N}\frac{s - z_{i}}{s - s_{i}},
\end{equation}
where the pole-zero ($s_{i}$-$z_{i}$) pairs are restricted as
\begin{equation}
\label{eq:zp-rule}
    0 \leqslant s_{1} < z_{1} < s_{2} < z_{2} <
        \cdots < s_{\scriptscriptstyle N} < z_{\scriptscriptstyle N} \leqslant 1.
\end{equation}
Therefore, the effective permittivity of GENZ metamaterial is simply determined by a series of
pole-zero pairs in contrast to Eq.~\eqref{eq:stack}.

%%%%%%%%%%%%%%%%%%%%%%%%%%%%
%\section{Structure Design}%
%%%%%%%%%%%%%%%%%%%%%%%%%%%%

Based on the foregoing text, the design of the broadband GENZ metamaterial can be simply
arranged as follows:
\begin{enumerate}
    \item
        Determine a proper operating frequency range, in which not only the real part of
        effective permittivity of GENZ metamaterial can be near-zero,
        but also the loss (imaginary part of effective permittivity of GENZ metamaterial) can be effectively
        reduced by gains.
    \item
        Determine a series of pole-zero pairs in order to make the real part of effective permittivity
        of GENZ metamaterial be near-zero in the operating frequency range.
    \item
        Determine the volume fractions and thicknesses with respect to the pole-zero pairs via an inverse problem,
        in order to obtain the physical structure of GENZ metamaterial.
\end{enumerate}
In the following, these steps are introduced in detail, and an example is depicted in Fig.~\ref{fig:fig2}.

%%%%---------------------------------%%%
%\subsection{Operating Frequency Range}%
%%%%---------------------------------%%%

In order to reduce the loss,
the operating frequency range is dominated by the effective permittivity
of gain shell and the $s$-parameter.
The imaginary part of effective permittivity of gain shell dominates the
effect of gain molecules.
Therefore, it is proper to locate the operating frequency range between
the emission frequencies ($\omega_{01},\,\omega_{02}$) to make a good use
of the gain molecules.
However, the dramatic variations of the effective permittivity of gain shell
at the emission frequencies should be avoid.
Therefore, the density factors ($\kappa_{1},\,\kappa_{2}$) should be carefully
chosen to make the effective permittivity varies smoothly between the emission
frequencies, just as depicted in Fig.~\ref{fig:fig2}\blue{(a)}.

On the other hand, the imaginary part of $s$-parameter should also be considered.
Just as indicated in Fig.~\ref{fig:fig2}\blue{(b)}, it is clear that between the emission
frequencies, the imaginary part of $s$-parameter has two zero points ($\omega_{c1},\,\omega_{c2}$).
Therefore, the final operating frequency range, denoted as ($\omega_{a},\,\omega_{b}$),
should be localized between the two zero points.

%%%%-----------------------%%%
%\subsection{Zeros and Poles}%
%%%%-----------------------%%%

The second step is to determine a series of pole-zero pairs with respect
to the operating frequency range, and it can be finished as follows:

\begin{enumerate}
\item[(1).]
Determine the initial zeros and the corresponding poles.
    \begin{enumerate}
    \item[\rmnum{1}).]
        This step starts with the determination of zeros.
        According to Eqs.~\eqref{eq:s-paramter}, \eqref{eq:milton}, and \eqref{eq:zp-rule},
        it is clear that all zeros should be real functions of frequency.
        Therefore, it is simple to define the zeros as the real part of $s$-parameter
        \begin{equation}
        \label{eq:i-zero}
            z_{i} = z_{i}(\omega_{i}) = \Re(s(\omega_{i})),
        \end{equation}
        where the frequency $\omega_{i}$ is determined as follows
        \begin{equation}
        \label{eq:z-f}
            \omega_{i} = \omega_{a}
                + (\omega_{b} - \omega_{a})(i-1)/(N-1)
                \quad (\text{for } i=1,2,\ldots,N),
        \end{equation}
        and named as zero-frequency.
    \item[\rmnum{2}).]
        Regarding Eq.~\eqref{eq:zp-rule}, the corresponding poles
        can be determined as follows:
        \begin{equation}
        \label{eq:pole}
        s_{i} =
        \begin{cases}
            0 & (\text{for } i=1), \\
            (z_{i-1}+z_{i})/2 & (\text{for } i=2,3,\ldots,N).
        \end{cases}
        \end{equation}
        It is worth noting that due to the simple mixing rule, i.e., Eq.~\eqref{eq:i-layer-s},
        the first pole should be zero.
    \end{enumerate}
Based on the definition of zeros and poles, the effective permittivity exactly equals zero
at all zeros and extends to infinity at all poles under lossless condition,
which is depicted in Fig.~\ref{fig:fig2}\blue{(c)} in orange.
However the result changes when loss is taken into account,
i.e., the dashed curves in Fig.~\ref{fig:fig2}\blue{(c)}.
Therefore, the initial zeros should be revised under lossy condition.

\item[(2).]
Zero-variation under lossy condition.

\hspace{1em}
To vary the initial zeros under lossy condition,
the following equation is considered
\begin{equation}
\label{eq:vz-eq}
    \Re\left[\varepsilon_{e}(s(\omega))\right]_{\omega=\omega_{i}}
        = \Re\left[\varepsilon_{2}
        \prod_{i=1}^{N}\frac{s(\omega) - z'_{i}}
        {s(\omega) - s_{i}}\right]_{\omega=\omega_{i}}= 0
        \quad (\text{for } i=1,2,\ldots,N),
\end{equation}
where the varied zeros are denoted as $z'_{i}$, and the
values of frequency $\omega_{i}$ and poles $s_{i}$ are
from Eqs.~\eqref{eq:z-f} and \eqref{eq:pole} respectively.
Note that it is difficult to solve this equation directly.
However, based on the values of initial zeros, i.e., Eq.~\eqref{eq:i-zero},
it is can be numerically solved.
After the variation, the real part of effective permittivity will
equals to zero at the varied zeros, which is indicated as solid
curve in Fig.~\ref{fig:fig2}\blue{(c)}.
\end{enumerate}

%%%%-----------------------%%%
%\subsection{Inverse Problem}%
%%%%-----------------------%%%

According to Eqs.~\eqref{eq:di-sum}, \eqref{eq:stack}, and \eqref{eq:milton},
the inverse problem reads
\begin{equation}
\label{eq:inverse}
    \sum_{i=1}^{N}\frac{f_{i}d_{i}}{s-f_{i}}
        = \prod_{i=1}^{N}\frac{s-s_{i}}{s-z'_{i}} - 1,
\end{equation}
with respect to the varied zeros $z'_{i}$ and the poles $s_{i}$. The
right hand side of Eq.~\eqref{eq:inverse} can be expanded into
\begin{equation}
\label{eq:inverse-2}
    \prod_{i=1}^{N}\frac{s-s_{i}}{s-z'_{i}} - 1
        = \sum_{i=1}^{N}\frac{Q_{i}}{s-z'_{i}}.
\end{equation}
Therefore, the volume fraction of metal-gain capsules in $i$th layer
is just $f_{i} = z'_{i}$, and the thickness of $i$th layer is $d_{i} = Q_{i}/z'_{i}$.
After all these steps, the physical structure of GENZ metamaterial is finally determined.
The detailed derivation of Eqs.~\eqref{eq:inverse} and \eqref{eq:inverse-2}
is given in the Appendix.

%%%%%%%%%%%%%%%%%%%%%%%%%%%%%%%%%
%\section{Example and Mechanism}%
%%%%%%%%%%%%%%%%%%%%%%%%%%%%%%%%%
%In this section, example and physical mechanism are demonstrated to clarify the design
%of broadband GENZ metamaterial.
%%%%---------------%%%
%\subsection{Example}%
%%%%---------------%%%
Figure~\ref{fig:fig2}\blue{(c,\,d)} displays a schematic example about the design.
Figure~\ref{fig:fig2}\blue{(c)} displays the variations of effective permittivity of GENZ metamaterial
with respect to frequency in different conditions.
In lossless condition, $\Re(\varepsilon_{e}(\omega))$ exactly equals zero
at every initial zero-frequency and extends to infinity at the frequency corresponding
to the poles.
When loss is taken into account, $\Re(\varepsilon_{e}(\omega))$
becomes continuous in the operating frequency range.
However, without zero-variation, it is not near-zero in the
operating frequency range.
In addition, a finite $\Im(\varepsilon_{e}(\omega))$ is also introduced
due to the loss.
In contrast, with the varied zeros, $\Re(\varepsilon_{e}(\omega))$
becomes near-zero in the operating frequency range.
Yet the varied zeros also effect $\Im(\varepsilon_{e}(\omega))$,
due to the Kramers-Kronig relations.

Figure~\ref{fig:fig2}\blue{(d)} offers a comparison between the results with and without gain.
Clearly, $\Im(\varepsilon_{e}(\omega))$ can be dramatically reduced by gains.
In this example the average reduction on the loss can be about 50\%.

Finally, it is worth noting that by increasing the number of layer,
reducing the width of operating frequency range,
or properly increasing the volume fraction of gain shell in the metal-gain capsule,
better result can be obtained.

%%%%------------------------------%%%
%\subsection{Cancellation Mechanism}%
%%%%------------------------------%%%

The zero-variation strategy is not only effective in designing GENZ metamterial,
but also reveals the physical mechanism of broadband ENZ phenomenon,
i.e., the cancellation mechanism in electric displacement.
Moreover, due to the Milton representation, the cancellation mechanism is
vividly indicated in the spectral space.

Figure~\ref{fig:fig3} takes a schematic example to explain the cancellation mechanism
in spectral space.
Consider a 2-layer stack, the effective permittivity of each layer
follows Eq.~\eqref{eq:i-layer-s},
and the volume fraction follows $f_{1}>f_{2}$ for convenience.
Initially, there is only one pole-zero pair in the spectral space for the first stack only,
i.e., $s_{1}=0$ and $z_{1}=f_{1}$,
according to Eqs.~\eqref{eq:i-layer-s} and \eqref{eq:milton}.
By adding the second layer, it amounts to add an additional pole-zero pair
($s_{2}$ and $z_{2}$) in the spectral space.
On the left hand side of the additional pole $s_{2}$, the polarization is out of phase
with the response of the pole $s_{1}$, thus yielding a zero $z_{2}$ between the poles.
Therefore, the out-phase polarization can cause the cancellation of electric
displacement, leading to a near-zero effective permittivity.
By properly arrange a series of pole-zero pairs, the broadband ENZ can be achieved.
Contrary to the physical space,
the spectral space has no concern with physical structure,
thus it offers a more clear picture about the cancellation mechanism.

%%%%%%%%%%%%%%%%%%%%%%%
%\section{Conclusions}%
%%%%%%%%%%%%%%%%%%%%%%%

In this work, a strategy based on the Milton representation of effective
permittivity is proposed to design the broadband GENZ metamaterial.
The application of Milton representation not only simplifies the design
procedure, but also reveals the physical mechanism dominating the broadband
ENZ phenomenon.
Meanwhile, the potential applications of this strategy in practical design
is also implied, since it is can be carried out numerically.

Furthermore, not only limited in the design of flat multi-layer broadband GENZ
metamaterial, the strategy also offers a fundament to design broadband GENZ
metamaterial in other geometrical structures, and material with other
permittivity in a wide frequency range.

%%%%%%%%%%%%%%%%%%%%
\section*{Appendix}%
%%%%%%%%%%%%%%%%%%%%

The derivation of Eqs.~\eqref{eq:inverse} starts with Eq.~\eqref{eq:stack},
which can be written as
\begin{equation*}
    \frac{\varepsilon_{2}}{\varepsilon_{e}}-1
        = \sum_{i=1}^{N}\frac{d_{i}}{1-f_{i}/s} - 1.
\end{equation*}
Regarding Eq.~\eqref{eq:di-sum}, the above equation can be rewritten as
\begin{equation*}
    \frac{\varepsilon_{2}}{\varepsilon_{e}}-1
        = \sum_{i=1}^{N}\left(\frac{d_{i}}{1-f_{i}/s} - d_{i}\right)
        = \sum_{i=1}^{N}\frac{f_{i}d_{i}}{s-f_{i}}.
\end{equation*}
On the other hand, Eq.~\eqref{eq:milton} gives
\begin{equation*}
    \frac{\varepsilon_{2}}{\varepsilon_{e}}-1
        = \prod_{i=1}^{N}\frac{s - s_{i}}{s - z_{i}} - 1.
\end{equation*}
Finally, with respect to the varied zeros $z'_{i}$, the inverse problem,
Eq.~\eqref{eq:inverse}, reads,
\begin{equation*}
    \sum_{i=1}^{N}\frac{f_{i}d_{i}}{s-f_{i}}
        = \prod_{i=1}^{N}\frac{s - s_{i}}{s - z'_{i}} - 1.
\end{equation*}
Regarding the derivation of Eq.~\eqref{eq:inverse-2}, it can be numerically carried out by
applying \emph{Mathematica}~8 of \emph{Wolfram Research}, or other similar computational softwares.

%%%%%%%%%%%%%%%%%%
%%% References %%%
%%%%%%%%%%%%%%%%%%

%%%%%%%%%%%%%%%%%%%%%%%%%%%
\newpage                  %
\section*{Figure Captions}%
%%%%%%%%%%%%%%%%%%%%%%%%%%%

\textbf{FIG.~1}. (Color on line)
The broadband GENZ metamaterial is a flat $N$-layer structure (a).
Each layer is composed by metal-gain capsules (b) doped in dielectric host.
The metal-gain capsule is a metallic core covered by a gain shell.
The permittivity of metallic core is denoted as $\varepsilon_{m}(\omega)$.
The gain shell is a shell of dielectric host doped with two kinds of gain molecules,
and its effective permittivity is denoted as $\varepsilon_{g}(\omega)$.
In addition, the effective permittivity of metal-gain capsule is denoted
as $\varepsilon_{1}(\omega)$, calculated by the simple mixing rule,
and the permittivity of dielectric host is set as $\varepsilon_{2}=\varepsilon_{r}$.
The thickness of $i$th layer is denoted as $d_{i}$, restricted by the summation rule.

\begin{figure}[htbp]
    \centering
    \includegraphics[width=8.0cm]{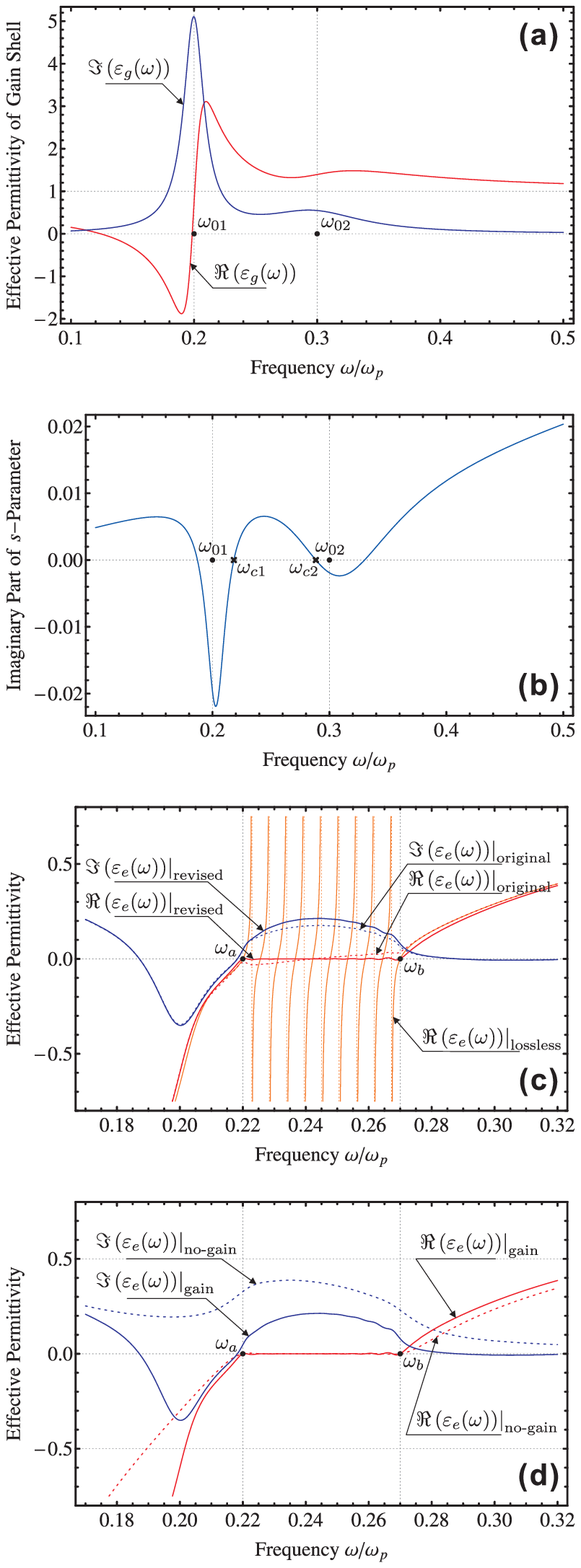}
    \caption{L.Sun and K.W.Yu}
    \label{fig:fig2}
\end{figure}

\vspace{5.0mm}
\textbf{FIG.~2}. (Color on line)
Figure~\ref{fig:fig2} indicates a schematic example about the GENZ metamaterial
design with the following parameters
(\rmnum{1}) high frequency permittivity $\varepsilon_{\scriptscriptstyle\infty}=1.0$,
(\rmnum{2}) damping factor $\gamma=0.02$,
(\rmnum{3}) permittivity of dielectric host $\varepsilon_{2}=\varepsilon_{r}=1.0$,
(\rmnum{4}) emission frequency $\omega_{01}=0.2$ and $\omega_{01}=0.3$,
(\rmnum{5}) emission frequency line-width $\sigma_{1}=0.02$ and $\sigma_{2}=0.1$,
(\rmnum{6}) density factor $\kappa_{1}=0.02$ and $\kappa_{2}=0.015$, and
(\rmnum{7}) volume fraction of gain shell $p=0.6$.
%%%
Figure~\ref{fig:fig2}\blue{(a,\,b)} displays the variation of effective permittivity
of gain shell and the imaginary part of $s$-parameter, which determine the location
of operating frequency range.
%%%
Figure~\ref{fig:fig2}\blue{(c)} denotes the effective permittivity of a 10-layer GENZ metamaterial
with respect to the operating frequency range $0.22\sim0.27$
determined by Fig.~\ref{fig:fig2}\blue{(a,\,b)}.
Under lossless condition, $\left.\Re(\varepsilon_{e}(\omega))\right|_{\text{lossless}}$ exactly equals
to zero at all zero-frequencies, but extends to infinity at the frequencies relating to the poles.
With respect to the initial zeros, $\left.\Re(\varepsilon_{e}(\omega))\right|_{\text{initial}}$ is not
near-zero in the operating frequency range, but increases slightly with respect to frequency.
After the zero-variation, $\left.\Re(\varepsilon_{e}(\omega))\right|_{\text{varied}}$ presents
a good result in the operating frequency rang.
Furthermore, it is clear that with respect to the near-zero result of
$\left.\Re(\varepsilon_{e}(\omega))\right|_{\text{varied}}$,
$\left.\Im(\varepsilon_{e}(\omega))\right|_{\text{varied}}$ is a little greater than
$\left.\Im(\varepsilon_{e}(\omega))\right|_{\text{initial}}$, due to the Kramers-Kronig relations.
%%%
Figure~\ref{fig:fig2}\blue{(d)} gives a compare between the results with and without gains.
Clearly, the gain does not affect the real part of effective permittivity too much
in the operating frequency range.
However, the loss is dramatically reduced by the gains and the average reduction is
about 50\% in this example.

\begin{figure}[htbp]
    \centering
    \includegraphics[width=8.0cm]{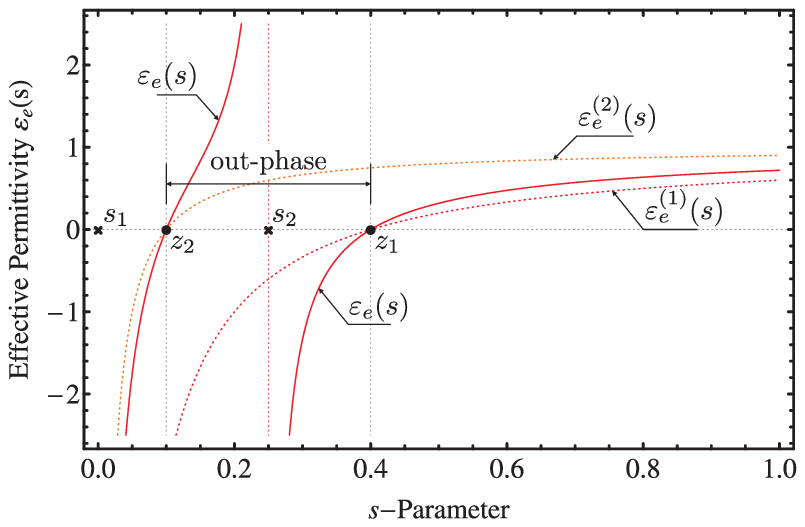}
    \caption{L.Sun and K.W.Yu}
    \label{fig:fig3}
\end{figure}

\vspace{5.0mm}
\textbf{FIG.~3}. (Color on line)
Figure~\ref{fig:fig3} gives a schematic explanation about the cacellation mechanism
in spectral space, which causes the broadband ENZ phenomenon.
It indicates the effective permittivity of a 2-layer stack (red-solid),
and the permittivity of each layer (red- and orange-dashed).
Consider only the first layer, there is only a pole-zero pair ($s_{1}$ and $z_{1}$).
By adding the second layer, it amounts to add an additional pole-zero pair ($s_{2}$ and $z_{2}$)
in the spectral space, thus yielding the out-phase polarizations by the sides
of the additional pole $s_{2}$.
The out-phase polarizations can cause the cancellation of electric displacement,
leading to the ENZ phenomenon.
Therefore, with a series properly arranged pole-zero pairs, the ENZ phenomenon
can occur in a wide frequency range.

%%%%%%%%%%%%%%%
%%% FIGURES %%%
%%%%%%%%%%%%%%%

%%% FIGURE 1 %%%
\newpage

%%% FIGURE 2 %%%
\newpage

%%% FIGURE 3 %%%
\newpage

%%% FIGURE 4 %%%
%\newpage
%\begin{figure}[htbp]
%    \centering
%    \includegraphics[width=7.50cm]{FIG-3-W-80mm.eps}
%    \caption{L.Sun and K.W.Yu}
%    \label{fig:fig4}
%\end{figure}

\end{document}